\documentclass[twocolumn,aps,prl,amsmath,amssymb,superscriptaddress,letterpaper]{revtex4}
\usepackage{times}
\usepackage{amsfonts}
\usepackage{mathrsfs}
\usepackage{graphicx}
\usepackage{dcolumn}
\usepackage{bm}
\usepackage{color}

\usepackage[colorlinks,bookmarks=false,citecolor=blue,linkcolor=red,urlcolor=blue]{hyperref}
\def\be{\begin{equation}}       \def\ee{\end{equation}}
\def\bea{\begin{eqnarray}}      \def\eea{\end{eqnarray}}

\begin{document}
\title{Robust $d_{x^2-y^2}$-wave superconductivity  of infinite-layer nickelates}

\author{Xianxin Wu}
\thanks{These authors equally contributed to the work.}
\email{xianxinwu@gmail.com}
\affiliation{Institut f\"{u}r Theoretische Physik und Astrophysik, Universit\"{a}t W\"{u}rzburg, Am Hubland Campus S\"{u}d, W\"{u}rzburg 97074, Germany}

\author{Domenico Di Sante}
\thanks{These authors equally contributed to the work.}
\affiliation{Institut f\"{u}r Theoretische Physik und Astrophysik, Universit\"{a}t W\"{u}rzburg, Am Hubland Campus S\"{u}d, W\"{u}rzburg 97074, Germany}

\author{Tilman Schwemmer}
\affiliation{Institut f\"{u}r Theoretische Physik und Astrophysik, Universit\"{a}t W\"{u}rzburg, Am Hubland Campus S\"{u}d, W\"{u}rzburg 97074, Germany}

\author{Werner Hanke}
\affiliation{Institut f\"{u}r Theoretische Physik und Astrophysik, Universit\"{a}t W\"{u}rzburg, Am Hubland Campus S\"{u}d, W\"{u}rzburg 97074, Germany}

\author{Harold Y. Hwang}
\affiliation{Stanford Institute for Materials and Energy Sciences, SLAC National Accelerator Laboratory, Menlo Park, California 94025, USA}
\affiliation{Department of Applied Physics, Stanford University, Stanford, California 94305, USA}

\author{Srinivas Raghu}
\email{sraghu@stanford.edu}
\affiliation{Stanford Institute for Materials and Energy Sciences, SLAC National Accelerator Laboratory, Menlo Park, California 94025, USA}
\affiliation{Department of Physics, Stanford University, Stanford, California 94305, USA}

\author{Ronny Thomale}
\email{rthomale@physik.uni-wuerzburg.de}
\affiliation{Institut f\"{u}r Theoretische Physik und Astrophysik, Universit\"{a}t W\"{u}rzburg, Am Hubland Campus S\"{u}d, W\"{u}rzburg 97074, Germany}

\date{\today}

\begin{abstract}
Motivated by the recent observation of superconductivity in strontium
doped NdNiO$_2$, we study the superconducting instabilities in this system from various vantage points.  Starting with first-principles
calculations, we construct two distinct tight-binding models, a simpler
single-orbital as well as a   three-orbital model, both of which capture
the key low energy degrees of freedom to varying degree of accuracy.  We
study superconductivity in both models using the random
phase approximation (RPA).  We then analyze the problem at stronger coupling,  and study the dominant pairing instability in the associated t-J model limit. In all instances, the dominant
pairing tendency is in the $d_{x^2-y^2}$ channel, analogous to the
cuprate superconductors.

\end{abstract}


\maketitle

\noindent {\it Introduction --}
The observation of superconductivity in the infinite-layer nickelate
Nd$_{1-x}$Sr$_x$NiO$_2$~\cite{Li2019}  resurrects some of the perennial
questions in the field of unconventional  superconductivity of the
cuprates and  related materials~\cite{Lee2006,Keimer2015}.  As nickel
substitutes for copper in this system, the low energy manifold consists
primarily of the Ni-O plane.  Therefore, we are invited to revisit
whether copper itself is important for the  superconductivity exhibited
by the
cuprates~\cite{Maeno1994,Chaloupka2008,Kim2016,Anisimov1999,Lee2004,PhysRevLett.103.016401}.
Furthermore, to date, magnetism has not been observed in the parent
NdNiO$_2$ compound~\cite{Hayward1999,Hayward2003}.  One may therefore
question the extent to which close proximity to  long-range
antiferrromagnetism is an essential  ingredient in cuprate
superconductivity.

\begin{figure}[b!]
\centering
\includegraphics[width=\columnwidth,angle=0,clip=true]{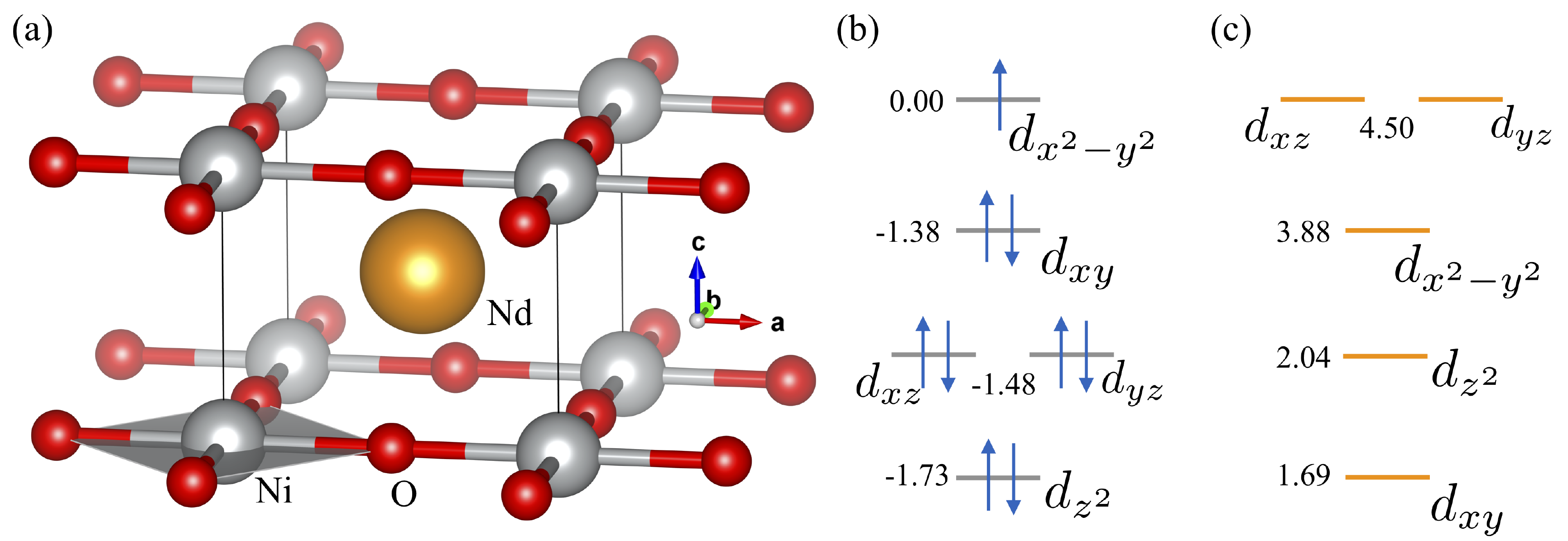}
\caption{(a) View of the crystal structure of NdNiO$_2$. Ni, O, and Nd atoms are
represented by grey, red and orange spheres. The planar coordination
in the NiO$_2$ is highlighted by a grey square. (b) The resulting
crystal field is characterized by a top $d_{x^2-y^2}$ orbital, which is
singly occupied in a $d^9$ electronic configuration. (c) Crystal field acting on the
formally empty Nd $d$ orbitals. In (b) and (c), the crystal field levels are given in eV,
with respect to the Ni $d_{x^2-y^2}$.}
\label{fig1}
\end{figure}

To help address these questions, we have studied superconductivity from
repulsive interactions in Nd$_{1-x}$Sr$_x$NiO$_2$, adopting both  weak- and strong-coupling approaches.
Starting with a first-principles study of NdNiO$_2$, and treating the
effects of strontium doping as a rigid shift to the chemical potential,
we have obtained tight-binding fits to the electronic structure.  As Ni
is isoelectronic to copper in this material, it has a $d^9$
configuration and the low energy physics is dominated by electrons in
the Ni-$d_{x^2-y^2}$ orbital (see Fig.~\ref{fig1}).  There is, however, an additional strong
hybridization with the $5d$ orbitals of the rare earth Nd element.  As a consequence, there is a non-zero
contribution to the low energy physics from the Nd $d_{z^2}$ and
$d_{xy}$ orbitals, which acts to introduce some distinction between this
system and the infinite layer cuprate material.

However, rather than speculating on the commonalities and differences
of the infinite-layer cuprate and nickelate, we have instead chosen to
study superconductivity in the nickelate material as a legitimate
problem in its own right, one that is independent from the cuprates.
The weak-coupling approach, while likely unreliable for normal state
properties, does tend to capture the primary property of interest,
namely the superconducting ground state itself and, in particular, the
symmetry of the superconducting order parameter.  We find robust
$d_{x^2-y^2}$-wave superconductivity within the weak-coupling approach.  We
have obtained this pairing symmetry in two distinct tight-binding fits
to the first-principles calculation, one which is a minimal one-orbital
model consisting of the Ni $d_{x^2-y^2}$ orbital, and a more
realistic 3-orbital model that includes the  $d_{z^2}$, $d_{xy}$
orbitals of the Nd atom.

In reality, however, the system is likely located at intermediate
coupling; it therefore becomes important to analyze the problem from
complementary limits.  With this in mind, we also analyze the t-J
model that results from the limit of strong Ni onsite interactions, and
study superconductivity in this model within a mean-field
approximation.  Such methods led to the conclusion of $d$-wave pairing
in the early days of cuprate physics~\cite{Kotliar1988}, and we arrive
at a similar conclusion in the present context.  We also show that
with the inclusion of the Nd electron pockets, $d_{x^2-y^2}$ pairing
stemming from the effective t-J model is only weakly affected.  While these electron pockets ultimately lead to metallic, rather than Mott insulating behavior in the parent compound, their impact on superconductivity appears to be rather weak.  The fact that all limits studied here result in $d_{x^2-y^2}$  pairing  underlies the
robustness of our conclusion.

This Rapid Communication is organized as follows.  At first, we present
the results of the first-principles computations, where we describe both
the minimal single-band and 3-band tight-binding fits to the electronic
structure.  We then proceed to show our results for the pairing symmetry both in
a random phase approximation (RPA) treatment of superconductivity from repulsive interactions, as
well as from the analysis of a t-J model description. Both
complementary studies are carried out in three dimensions (3D), corresponding to the infinite-layer limit.

\noindent {\it  First-principles analysis --} We performed first-principles
calculations within the framework of the density functional theory as
implemented in the Vienna ab initio simulation package VASP~\cite{vasp1,vasp2,paw}. The
generalized gradient approximation, as parametrized by the PBE-GGA
functional for the exchange-correlation potential, was used, by
expanding the Kohn-Sham wave functions into plane waves up to an energy
cutoff of 600 eV and sampling the Brillouin zone on an 16$\times$16$\times$16 regular
mesh~\cite{pbe}. The growth of NdNiO$_2$ on a SrTiO$_3$ substrate is simulated by
imposing an in-plane lattice constant $a = 3.91$\AA\, and relative relaxed
out-of-plane parameter $c = 3.37$~\AA~\cite{Li2019}. The extraction of the three-orbitals minimal model
used to investigate the superconducting tendencies of NdNiO$_2$ was based on the Wannier
functions formalism~\cite{wann90}.

\begin{figure}[t]
\centering
\includegraphics[width=\columnwidth,angle=0,clip=true]{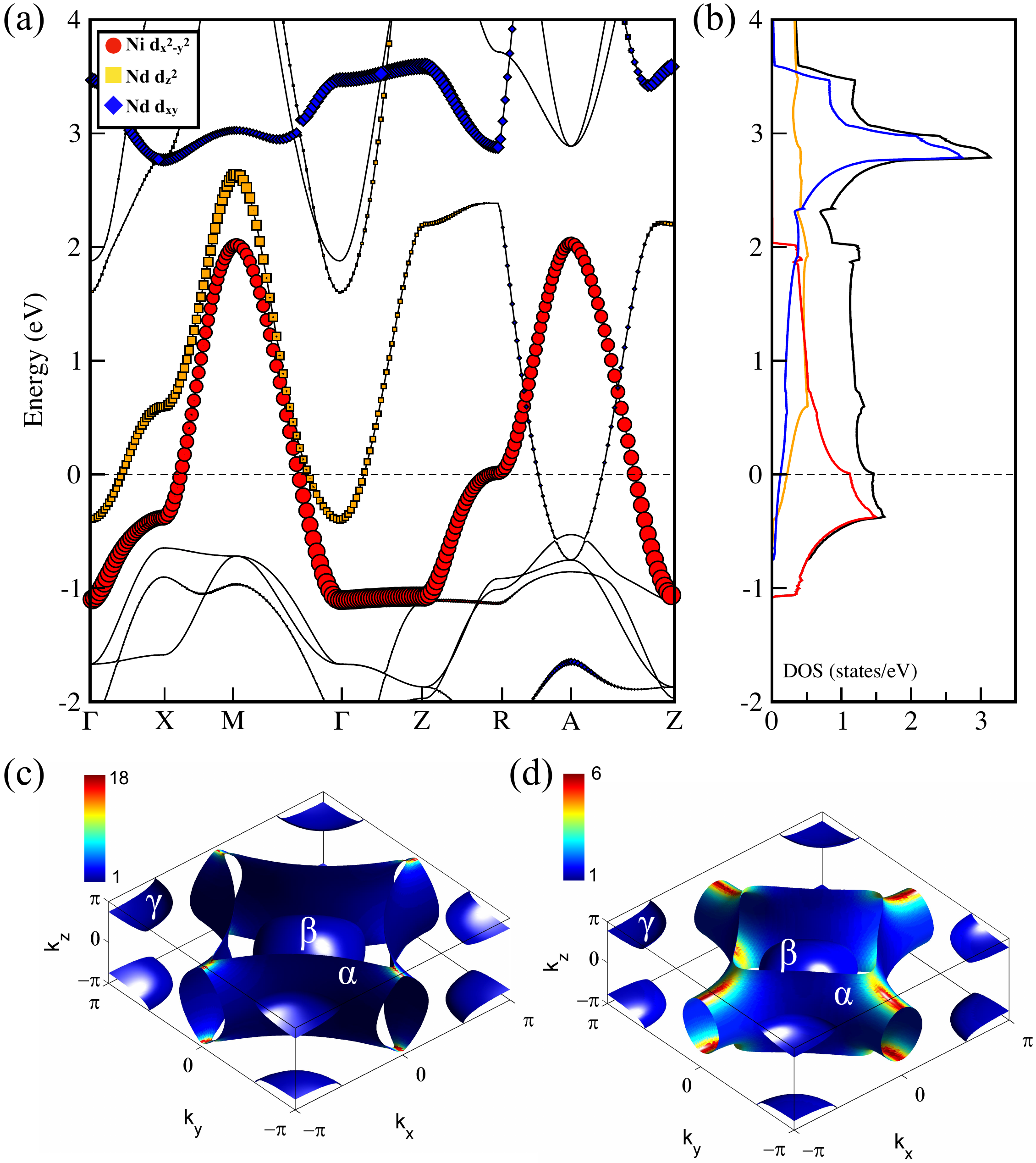}
\caption{First-principles band structure (a) and density of states (b) of NdNiO$_2$ with lattice
parameters forced by the commensuration to the SrTiO$_3$
substrate. The red (Ni $d_{x^2-y^2}$), yellow (Nd $d_{z^2}$), and blue
(Nd $d_{xy}$) symbols emphasize the
relevant orbitals that contribute to the low-energy description. In panel (b) the black curve refers to the sum
of the individual contributions.
Views of the Fermi surface of NdNiO$_2$ at (c) pristine
filling $(n = 1.0)$ and (d) upon 0.2 hole doping $(n=0.8)$. The color scale
reports the momentum dependence of the inverse Fermi velocity $(1/v_F({\textbf k}))$,
which is a measure of the DOS. The $\alpha$ Fermi surface displays a
van Hove feature evolving from the $k_z=0$ cut to the $k_z=\pi$ cut, where it changes from a hole pocket around the M point in the $k_z=0$ plane to an electron pocket around the Z point in the $k_z=\pi$ plane.}
\label{fig2}
\end{figure}

Fig.~\ref{fig2} shows the single-particle band structure
of NdNiO$_2$, along with the orbital contributions relevant for the low-energy model
description. Owing to a $d^9$ electronic configuration in a peculiar $+1$ oxidation
state for Ni, the crystal field imposed by the planar square coordination (Fig.~\ref{fig1}) results in
a high-lying nominally half-filled $d_{x^2-y^2}$ orbital (red dots), featuring a predominantly two-dimensional
character. Nonetheless, the delocalized and formally empty Nd $5d$ states reside fairly low
in energy, leading to a sizable hybridization with Ni $3d$ bands, and to the appearance
of electron pockets at the $\Gamma$ and ${\text A}=(\pi/a,\pi/a,\pi/c)$
(see panel Fig.~\ref{fig2}(a)) points. Such pockets mainly display  Nd $d_{z^2}$ (yellow squares)
and $d_{xy}$ (blue diamonds) orbital contributions, respectively, and determine a concomitant
self-doping of the large hole-like Ni $d_{x^2-y^2}$ Fermi surface.


Having established the contribution of the relevant orbitals to the low-energy
physics of NdNiO$_2$, we consider a three-orbital tight-binding (TB) model which includes
long range hopping terms. We introduce the operator
$\psi^\dag_{\textbf{k}\sigma}=[c^\dag_{1\sigma}(\textbf{k}),c^\dag_{2\sigma}(\textbf{k}),c^\dag_{3\sigma}(\textbf{k})]$,
where $c^\dag_{\alpha\sigma}(\textbf{k})$ is a fermionic creation
operator with  $\sigma$ and $\alpha$ denoting spin and orbital indices, respectively. The orbital index $\alpha=1,2,3$ represents the Nd $d_{z^2}$ for 1, the Nd $d_{xy}$ for 2 and the Ni $d_{x^2-y^2}$ for 3. The tight-binding Hamiltonian can be written as
 \begin{eqnarray}
 H_{\text{TB}}=\sum_{\textbf{k}\sigma}\psi^\dag_{\textbf{k}\sigma}h(\textbf{k})\psi_{\textbf{k}\sigma}, \label{tbmodel}
 \end{eqnarray}
where $h(\bm{k})$ is given in~\cite{supp}, along with the
corresponding parameters extracted from a downfolding of the
first-principles band structure onto a set of localized Wannier
functions. With the above parameters, the obtained band structure fits
are given in~\cite{supp} and reach a good agreement between DFT and
the effective TB bands. Near the Fermi level, the DOS is dominantly
attributed to the Ni $d_{x^2-y^2}$ orbital, as shown in
Fig.~\ref{fig2}(b). Further considering the relatively weak
interaction effects in the 5d orbitals of Nd, the dominant correlation
effects must derive from the 3d $d_{x^2-y^2}$ orbital of Ni in
NdNiO$_2$.  These conclusions are consistent with
previous~\cite{Anisimov1999,Lee2004} as well as concurrent~\cite{Botana2019} first principles calculations of this system.

The resulting 3D Fermi surfaces are shown in Fig.~\ref{fig2}(c) and (d) for the fillings $n=1.0$ and $n=0.8$. For the former case, there is an
almost cylindrical, non-dispersive in $k_z$, hole-like pocket $\alpha$, and two small
electron-like pockets $\beta$ and $\gamma$ around the $\Gamma$ and ${\text A}$ points, respectively.
With 0.2 hole doping, the electron Fermi surfaces shrink.
For the hole pocket, van Hove singularities are reached near the $k_z=\pi$ plane,
and its density of states increases considerably along with enhanced
nesting, as shown by the red curve in Fig.~\ref{fig2}(b). The
three-dimensional character of the obtained Fermiology is an essential
distinguishing aspect from the infinite-layer cuprates.
Within a weak-coupling
framework of superconductivity, such enhancement of the density of
states available for pairing as obtained for hole doping in the nickelates
 typically results in a concomitant increase of the superconducting
temperature. The largest contribution to the density of states arises from the large
non-$k_z$-dispersive Ni $d_{x^2-y^2}$ pocket, suggesting that, to some
approximation, it likely plays a significant role in the superconducting transition.

\noindent {\it Weak-coupling Analysis --}
In order to investigate the pairing symmetry of NdNiO$_2$, we first
consider a weak-coupling limit of the problem. Weak coupling can
either be interpreted as the strictly analytically controlled
perturbative limit of interactions~\cite{Kohn1965,Raghu2010, Cho2013}
or, in a less restrictive meaning, as the itinerant electronic limit in which a
diagrammatic, e.g. RPA treatment of interactions is adopted starting from the
bare or effective electronic band structure.
Such or related approaches, while less controlled than the strictly
perturbative weak coupling limit, are more physically motivated, and
result in qualitatively similar conclusions for pairing strengths in the
system. They have enjoyed significant phenomenological success in
describing unconventional superconductivity~\cite{Scalapino2012}.  For
instance, in both the weak coupling and RPA treatments of the single
band Hubbard model, the dominant pairing tendency near half-filling is
in the $d_{x^2-y^2}$ channel~\cite{Scalapino1986, Miyake1986}. A
perturbative combined diagrammatic inclusion of particle-particle and particle-hole
contributions could be reached by the employment of functional
renormalization group~\cite{RevModPhys.84.299,doi:10.1080/00018732.2013.862020} in order to further sophisticate the RPA treatment. For the
case at hand, however, the absence of magnetic
order combined with the enhanced feasibility in treating
three-dimensional band structures render the RPA approach most
preferable at this stage of our weak coupling analysis.

In our RPA calculations, we consider onsite Hubbard intra- and
inter-orbital repulsion, Hund's coupling as well as pair hopping interactions,
\begin{eqnarray}
H_{\text{int}}&=&U_{\text{Ni}}\sum_{i}n_{i3\uparrow}n_{i3\downarrow}+U_{\text{Nd}}\sum_{i\mu}n_{i\mu\uparrow}n_{i\mu\downarrow}\nonumber\\
&+&U'_{\text{Nd}}\sum_{i,\mu<\nu}n_{i\mu}n_{i\nu}+J_{\text{Nd}}\sum_{i,\mu<\nu,\sigma\sigma'}c^{\dag}_{i\mu\sigma}c^{\dag}_{i\nu\sigma'}c_{i\mu\sigma'}c_{i\nu\sigma}\nonumber\\
&+&J'_{\text{Nd}}\sum_{i,\mu\neq\nu}c^{\dag}_{i\mu\uparrow}c^{\dag}_{i\mu\downarrow}c_{i\nu\downarrow}c_{i\nu\uparrow}
\label{interaction2}
\end{eqnarray}
where $n_{i\alpha}=n_{\alpha\uparrow}+n_{\alpha\downarrow}$,
$\mu,\nu=1,2$, $U_{\text{Ni}}$ is the Coulomb repulsion for the Ni site,
thus acting on the third orbital in the notation
of Eq.~\eqref{tbmodel}. $U_{\text{Nd}}$, $U'_{\text{Nd}}$, $J_{\text{Nd}}$ and $J'_{\text{Nd}}$
represent the onsite intra- and inter-orbital repulsion and the onsite
Hund's coupling and pair-hopping terms for the Nd site,
respectively. For simplicity, we have chosen the same value of $U$ for
both the Nd and Ni $d$-orbitals. This initial choice does not
fundamentally affect the effective BCS coupling that arises at order
$U^2$, as it is weighted by the susceptibility of each orbital. As
such, the BCS couplings at low energies systematically are significantly different for Nd and Ni electrons. We use the Kanamori relations $U_{\text{Nd}}=U'_{\text{Nd}}+2J_{\text{Nd}}$ and $J_{\text{Nd}}=J'_{\text{Nd}}$.

\begin{figure}[tb]
\centerline{\includegraphics[width=1.0\columnwidth]{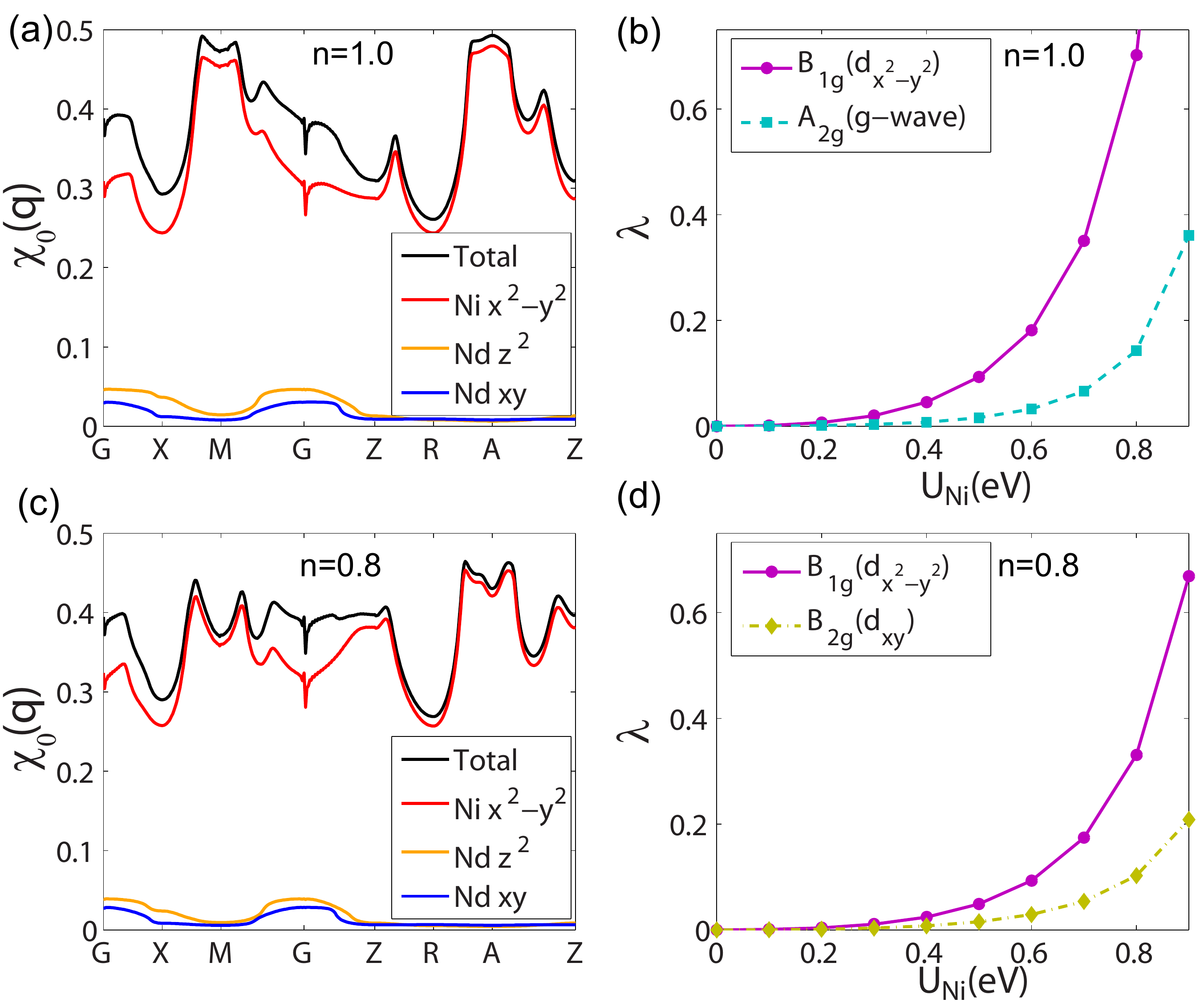}}
\caption{Bare susceptibility (left panels) and pairing eigenvalues as a function of the interaction $U_{\text{Ni}}$ (right panels) for
electron filling $n = 1.0$ (top row) and $n = 0.8$ (bottom row), respectively. Here we adopt $U_{\text{Ni}}=U_{\text{Nd}}$ and $J_{\text{Nd}}/U_{\text{Nd}}=0.15$.}
\label{sus}
\end{figure}

Fig.~\ref{sus} displays the bare susceptibilities for $n=1.0$
and $n=0.8$ filling, respectively. In both cases, similar to cuprates, the dominant peaks
are located around the M and A points, indicating intrinsic
antiferromagnetic fluctuations. These peaks  get significantly enhanced
upon including interactions at the RPA level. The prominent features in the
orbital-resolved susceptibility
are that the peaks around M
and A are dominantly attributed to the Ni $d_{x^2-y^2}$ orbital while the
contribution of Nd $d_{xy}$ and $d_{z^2}$ reaches its maximum around
$\Gamma$. Based on the analysis of the susceptibility, the $d_{x^2-y^2}$
band will play the dominant role in promoting correlation phenomena,
including superconductivity and, if commensurate filling of this band
were reached, possible magnetic ordering.

As a systematic methodological feature, when the interaction is
greater than a critical value $U_c$ (1.1 eV in our case), the spin
susceptibility within RPA diverges and
indicates a spin density wave (SDW) instability. Note that as a matter
of principle, $U_c$ should be interpreted as a phenomenological
parameter that does not allow an immediate quantitative connection
with the bare unrenormalized interaction strength. Below $U_c$,
superconductivity emerges triggered by spin fluctuations. We perform RPA calculations to study the possible pairing
symmetries within a $40\times40\times20$ $k$ mesh, energy window $\Delta E=0.02$ eV around the Fermi level and
inverse temperature $\beta=50$ eV$^{-1}$, and have checked the convergence of pairing strength with respect to the $k$ mesh and $\Delta E$. With the above parameters, the
numbers of the representative momentum points on the Fermi surface are
1038 and 1088 for $n=1.0$ and $n=0.8$, respectively. From the
susceptibility, we can expect the dominant pairing state to be
$d_{x^2-y^2}$-wave (more details on the RPA are provided in the
supplement~\cite{supp}). As the effective low-energy interaction parameters
remain as of yet largely undetermined for infinite layer nickelates,
we have performed our RPA calculations within a large region of
parameter space, and consistently found the dominant pairing
to be unchanged, and, in particular, largely insensitive to the bare
initial value $U_{\text{Nd}}$. The obtained pairing
eigenvalues as a function of interaction strength $U$ for $n=1.0$ and $n=0.8$ are
 displayed in Fig.~\ref{sus}. We find that $d_{x^2-y^2}$ pairing
state is dominant, and that the gap functions are
 considerably smaller on the two small spherical Fermi surfaces. This
 is consistent with the fact that both the dominant density of states and pairing interactions reside on the Ni $d_{x^2-y^2}$ orbital.


\begin{figure}[tb]
\centerline{\includegraphics[width=1.0\columnwidth]{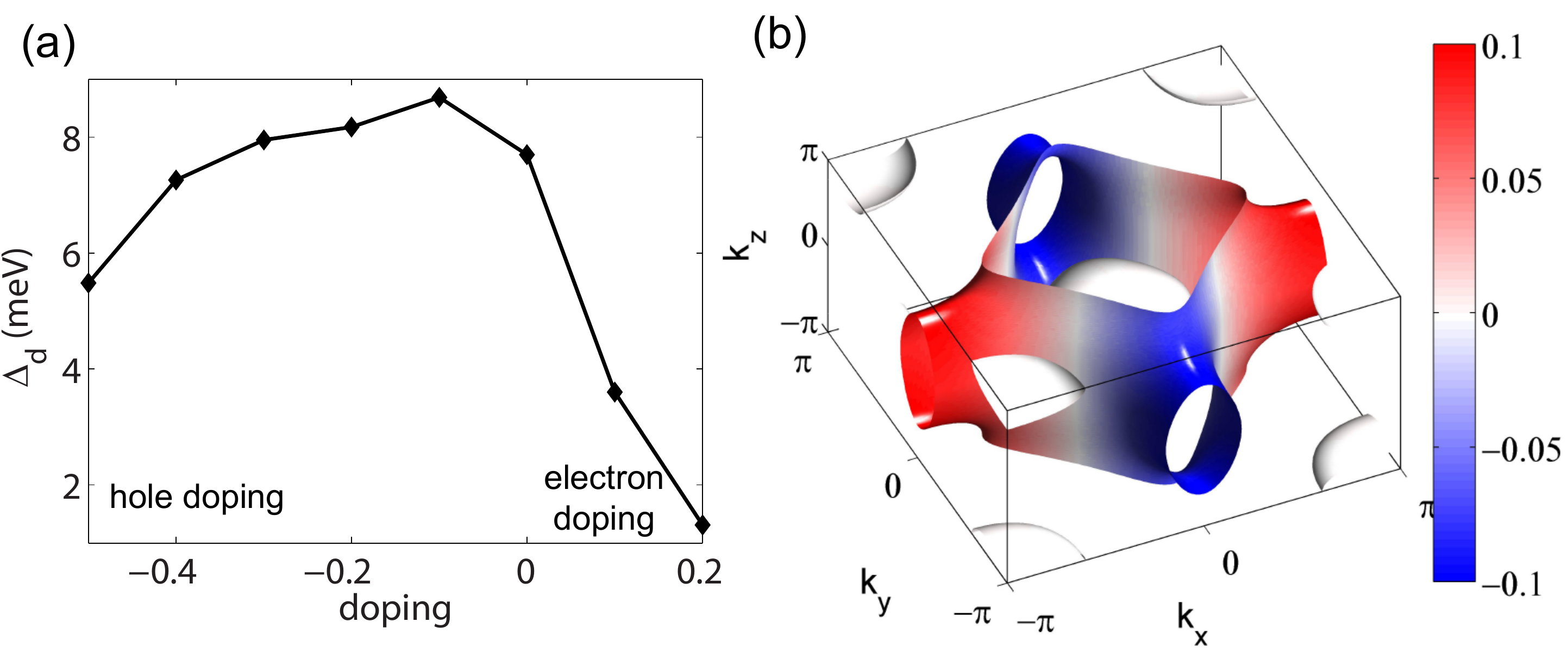}}
\caption{ (a) The $d_{x^2-y^2}$-wave gap as a function doping with
  $J_1=J_2=0.1$ eV. Postive (negative) values relate to electron
  (hole) doping. (b) Superconducting gap for $d_{x^2-y^2}$-wave
  pairing. In the calculations, a $k$ mesh of $100\times100\times50$
  has been adopted.}
\label{tjresult}
\end{figure}

\noindent {\it Pairing in the t-J model --} Similar to cuprates, the
nickelates represent an intermediately coupled system, and it becomes
important to ``triangulate" the pairing problem from various limits to
see if our conclusions are indeed robust. From a strong coupling
perspective, the cuprates have been addressed within an effective t-J
model which is either obtained from the Gutzwiller projection of a
single band Hubbard model or the low-energy perturbative description
of the three-band Hubbard model involving the Cu $d_{x^2-y^2}$ and the planar O
$p_{x,y}$ orbitals~\cite{Zhang1988}. As of yet, it is unclear whether the
charge transfer gap in the nickelates~\cite{Botana2019} allows for a Zhang-Rice
type Ni-O singlet complex of two holes as a suitable effective
description. Still, approaching the nickelates from a related angle, we adopt the t-J model reduced to the Ni
$d_{x^2-y^2}$ orbital. In doing so, we describe a strong coupling limit of
the doped Ni $d_{x^2-y^2}$  band by constraining ourselves to the in-plane and out-of-plane antiferromagnetic couplings
between the Ni spins:
\begin{eqnarray}
H_{J}=\sum_{\langle ij\rangle}J_{ij}(\mathbf{S}_{i3}\mathbf{S}_{j3}-\frac{1}{4}n_{i3}n_{j3}),
\end{eqnarray}
where
$\bm{S}_{i3}=\frac{1}{2}c_{i3\sigma}^{\dagger}\bm{\sigma}_{\sigma\sigma'}c_{i3\sigma'}$
is the local spin operator and $n_{i3}$ is the local density
operator for the Ni $d_{x^2-y^2}$ orbital. $\langle ij\rangle$ denotes
the in-plane and out-of-plane nearest neighbor (NN). The in-plane
coupling is $J_x=J_y=J_1$ and the out-of-plane coupling is $J_2$. We
investigate the pairing state for an extended range of doping levels. We
simplify the analysis by relaxing the double-occupancy constraint on this t-J model, perform a mean-field
decoupling, and solve the self-consistent gap equations~\cite{supp}. We find that $d_{x^2-y^2}$ pairing is
always the dominant order within extended parameter ranges of $J_1$
and $J_2$, and that $J_2$ has a negligible effect on the
gap. Fig.~\ref{tjresult}(a) shows the representative superconducting gap of the
$d_{x^2-y^2}$ pairing as a function of doping with $J_1=J_2=0.1$ eV. We
find that there is a superconducting dome and the gap reaches the
maximum upon 0.1 hole doping. Electron doping, by reducing the
contribution of the Ni $d_{x^2-y^2}$ orbital, will significantly
suppress superconductivity. Instead, beyond optimal doping, further
hole doping will only slightly suppress the superconducting gap,
implying to expect an extended $T_c$ dome feature on the hole doped
side. The 3D gap function of the obtained $d_{x^2-y^2}$-wave pairing
is displayed in Fig.~\ref{tjresult}(b) at 0.2 hole doping, where the
gaps on the spherical Fermi surfaces from Nd atoms almost vanish. Our
findings from a t-J model analysis thus are consistent with our weak-coupling analysis.

\noindent {\it Discussion --}
We have studied the infinite-layer nickelate NdNiO$_2$ and have found
that the dominant pairing instability is in the $d_{x^2-y^2}$ channel,
which places this system in close analogy with cuprate superconductors.
As a consequence of the pairing symmetry, we expect nodes on the Fermi
surface, the evidence for which can be found in low temperature heat
capacity~\cite{Moler1995}, penetration depth~\cite{Hardy1993}
measurements, quasiparticle interference studies~\cite{Hoffmann2002},
and more directly, from phase-sensitive
studies~\cite{Vanharlingen1995,Tsuei2000}.

In the future, it will be interesting to study the role of the Nd itinerant electrons in conjunction with
the local moments of the Ni sites.   It is thus tempting to invoke the analogy with heavy fermion systems, and to view the physics of the infinite layer nickelate from the vantage point of the Kondo
lattice.  In this
context, it is reasonable to presume that the effect of strontium doping
involves more complex phenomena than a simple rigid shift of the Fermi
level. Furthermore, even though an electronically mediated pairing
mechanism may appear likely judging from the current experimental
evidence, the impact of electron-phonon coupling will be vital to
gaining a deeper understanding of the material~\cite{AritaPRB}. We wish to pursue such questions in future studies.

\noindent {\it Acknowledgments - }
The work in W\"urzburg is funded by the Deutsche
Forschungsgemeinschaft (DFG, German Research Foundation) through
Project-ID 258499086 - SFB 1170 and through the W\"urzburg-Dresden Cluster of Excellence on Complexity and Topology in Quantum Matter --\textit{ct.qmat} Project-ID 39085490 - EXC 2147. H.Y.H. and S.R. are supported by the  DOE Office of
Basic Energy Sciences, contract DEAC02-76SF00515. We gratefully acknowledge the
Gauss Centre for Supercomputing e.V. (www.gauss-centre.eu) for funding
this project by providing computing time on the GCS Supercomputer
SuperMUC at Leibniz Supercomputing Centre (www.lrz.de).

\noindent {\it Note added -}
After completion of our work, we learned about an independent study of
electronic structure and pairing instabilities in the infinite layer
nickelate (Ref. ~\cite{Hirofumi2019}).  This study makes use of a
variant of the RPA method (FLEX) and finds similar conclusions for pairing.


\renewcommand{\theequation}{S\arabic{equation}}
\renewcommand{\thefigure}{S\arabic{figure}}
\renewcommand{\bibnumfmt}[1]{[S#1]}
\renewcommand{\citenumfont}[1]{S#1}

\clearpage
\appendix
\begin{widetext}

  \section{DFT bands and Tight-binding model}
Fig.\ref{dftband} shows the DFT band structure with a large energy window and the band structures in $k_z=0$ and $k_z=\pi$ plane. In the $k_z=\pi$ plane, the van Hove singularity point at R is extremely close to the Fermi level. To reproduce the DFT band structure we construct a tight-binding model, which was given the main text. The adopted three-band tight binding Hamiltonian is given in Eq.1. The matrix elements in the Hamiltonian $h(\textbf{k})$ matrix are given by,
\begin{eqnarray}
\label{caas_tb}
h_{11}(\bm{k})&=&\epsilon_1-\mu+2t^x_{11}(cosk_x+cosk_y)+4t^{xy}_{11}cosk_xcosk_y+2t^{xx}_{11}(cos2k_x+cos2k_y)+4t^{xxy}_{11}(cos2k_xcosk_y+cosk_xcos2k_y)\nonumber\\
&&+2t^z_{11}cosk_z+2t^{zz}_{11}cos2k_z+4t^{xz}_{11}cosk_z(cosk_x+cosk_y)+8t^{xyz}_{11}cosk_xcosk_ycosk_z\\
h_{22}(\bm{k})&=&\epsilon_2-\mu+2t^x_{22}(cosk_x+cosk_y)+4t^{xy}_{22}cosk_xcosk_y+2t^{xx}_{22}(cos2k_x+cos2k_y)+4t^{xxy}_{22}(cos2k_xcosk_y+cosk_xcos2k_y)\nonumber\\
&&+2t^z_{22}cosk_z+2t^{zz}_{22}cos2k_z+4t^{xz}_{22}cosk_z(cosk_x+cosk_y)+8t^{xyz}_{22}cosk_xcosk_ycosk_z\nonumber\\
&&+8t^{xxz}_{22}cosk_z(cos2k_x+cos2k_y)+8t^{xxyz}_{22}cosk_z(cos2k_xcosk_y+cosk_xcos2k_y)\nonumber\\
&&+8t^{xxyz}_{22}cosk_z(cos2k_xcosk_y+cosk_xcos2k_y)+2t^{xxx}_{22}(cos3k_x+cos3k_y)\\
h_{33}(\bm{k})&=&\epsilon_3-\mu+2t^x_{33}(cosk_x+cosk_y)+4t^{xy}_{33}cosk_xcosk_y+2t^{xx}_{33}(cos2k_x+cos2k_y)+4t^{xxy}_{33}(cos2k_xcosk_y+cosk_xcos2k_y)\nonumber\\
&&+4t^{xxyy}_{33}cos2k_xcos2k_y+2t^z_{33}cosk_z+2t^{zz}_{33}cos2k_z+4t^{xz}_{33}cosk_z(cosk_x+cosk_y)+8t^{xyz}_{33}cosk_xcosk_ycosk_z\\
h_{12}(\bm{k})&=&-4t^{xy}_{12}sink_xsink_y-8t^{xyz}_{12}sink_xsink_ycosk_z-8t^{xyzz}_{12}sink_xsink_ycos2k_z\\
h_{13}(\bm{k})&=&-8t^{xxy}_{13}cosk_z/2(sin3k_x/2sink_y/2-sink_x/2sin3k_y/2)\\
h_{23}(\bm{k})&=&8t^{xxy}_{23}cosk_z/2(cos3k_x/2cosk_y/2-cosk_x/2cos3k_y/2)
\end{eqnarray}
The corresponding tight binding parameters are specified in unit of $eV$ as,
\begin{eqnarray}
&&  \epsilon_1=8.9506,\quad \epsilon_2=9.0277,\quad \epsilon_3=6.8979,\\
&& t^x_{11}=-0.3870, \quad t^{xy}_{11}=0,\quad t^{xx}_{11}=0.034,\quad t^z_{11}=-0.8591, \quad t^{xz}_{11}=0.0107,\quad t^{xyz}_{11}=0.025,\quad t^{zz}_{11}=0.0904,\\
&& t^x_{22}=0.3202, \quad t^{xy}_{22}=-0.0467,\quad t^{xx}_{11}=0.0367,\quad t^z_{11}=0.3216, \quad t^{xz}_{22}=-0.1438,\quad t^{xyz}_{22}=0.0496,\quad t^{zz}_{22}=-0.0327,\nonumber\\
&&t^{xxz}_{22}=-0.0209,\quad t^{xxy}_{22}=-0.0198,\quad t^{xxyz}_{22}=0.0164,\quad t^{xxx}_{22}=0.012\\
&&t^{xy}_{12}=0.0798, \quad t^{xyz}_{12}=-0.0669,\quad t^{xyzz}=0.0094,\\
&& t^x_{33}=-0.3761, \quad t^{xy}_{33}=0.0844,\quad t^{xx}_{33}=-0.0414,\quad t^{xyy}_{33}=-0.0043,
\quad t^{xxyy}_{33}=0.003,\quad t^z_{33}=-0.0368, \quad t^{xz}_{33}=-0.0019,\nonumber\\
&& \quad t^{xyz}_{33}=0.0117,\quad t^{zz}_{33}=0.008,\\
&& t^{xxy}_{13}=0.0219,\quad t^{xxy}_{23}=-0.0139.
\label{hopping}
\end{eqnarray}

With $\mu= 6.5814$ eV, the occupation number is 1.0. With 0.2 hole doping, the corresponding chemical potential is $\mu=6.4614$ eV. The band structures from the tight-binding model with the above parameters are displayed in Fig.\ref{tbband}, which are in good agreement with DFT calculations.

\begin{figure}[t]
\centering
\includegraphics[width=\columnwidth,angle=0,clip=true]{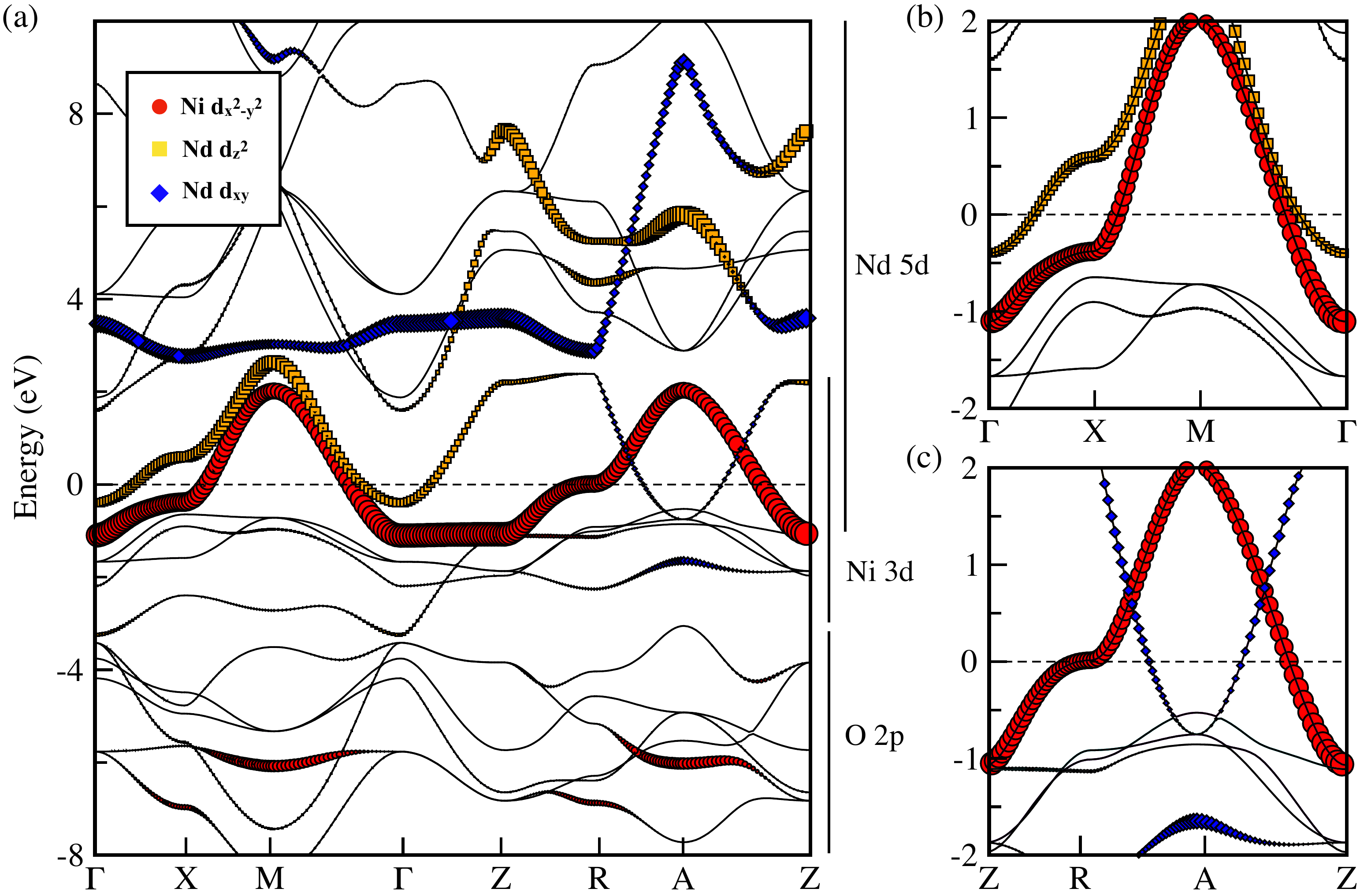}
\caption{(a) First-principles band structure of NdNiO$_2$ with lattice
parameters forced by the commensuration to the SrTiO$_3$ substrate. The
lateral bars qualitatively identify the range of energy extension of the
relative atomic contributions. A sizable hybridization between Ni $3d$ and
Nd $5d$ states is expected close the Fermi level. The red, yellow and blue symbols emphasize the
relevant orbitals that contribute to the low-energy description. (b-c) Enlarged
views of the band structure around the Fermi level at (b) $k_z = 0$ and (c)
$k_z = \pi/c$, respectively}
\label{dftband}
\end{figure}

\begin{figure}[tb]
\centerline{\includegraphics[width=0.6\columnwidth]{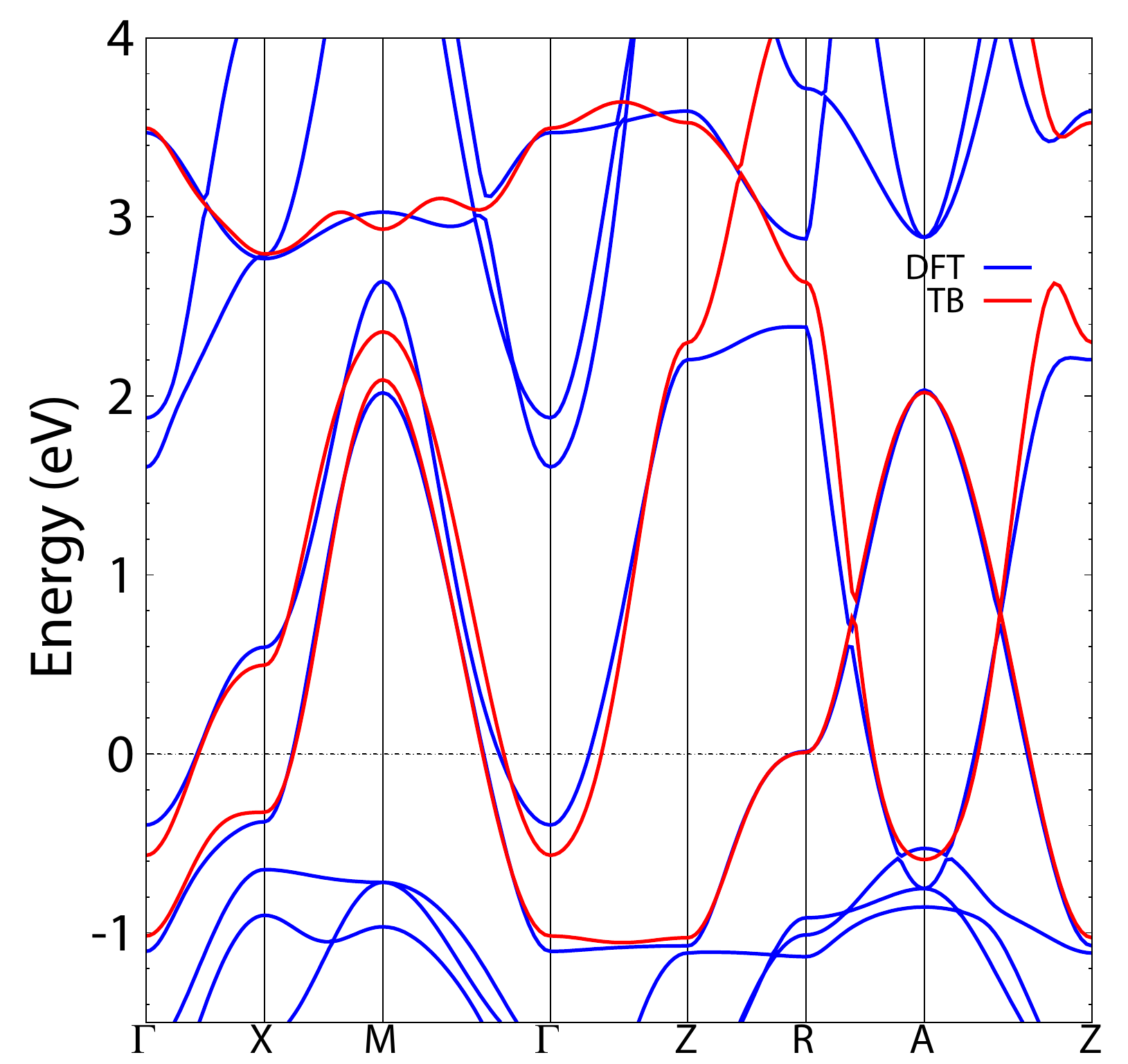}}
\caption{ Band structures from DFT (blue lines) and tight-binding model (red lines). }
\label{tbband}
\end{figure}

\section{weak-coupling limit: RPA approach} \label{rpamethod}

The adopted interactions are given in the main text. The bare susceptibility is define as,
\begin{eqnarray}
\chi^{0}_{l_1l_2l_3l_4}(\bm{q},\tau)=\frac{1}{N}\sum_{\bm{k}\bm{k}'}\langle T_{\tau} c^{\dag}_{l_3\sigma}(\bm{k}+\bm{q},\tau)c_{l_4\sigma}(\bm{k},\tau)c^{\dag}_{l_2\sigma}(\bm{k}'-\bm{q},0)c_{l_1\sigma}(\bm{k}',0) \rangle_0 .
\end{eqnarray}
where $l_i$ is the orbital indices. The bare susceptibility in momentum-frequency is,
\begin{eqnarray}
\chi^0_{l_1l_2l_3l_4}(\bm{q},i\omega_n)\!\!=\!\!-\frac{1}{N}\!\!\sum_{k\mu\nu}a^{l_4}_\mu(\bm{k})a^{l_2*}_{\mu}(\bm{k}) a^{l_1}_\nu(\bm{k}+\bm{q})a^{l_3*}_{\nu}(\bm{k}+\bm{q})\frac{n_F(E_{\mu}(\bm{k}))-n_F(E_{\nu}(\bm{k}+\bm{q}))}{i\omega_n+E_{\mu}(\bm{k})-E_{\nu}(\bm{k}+\bm{q})}.
\end{eqnarray}
where $\mu/\nu$ is the band index, $n_F(\epsilon)$ is the Fermi distribution function, $a^{l_i}_\mu(\bm{k})$ is the $l_i$-th component of the eigenvector for band $\mu$ resulting from the diagonalization of the tight-binding Hamiltonian $H_{TB}$ and $E_{\mu}(\bf{k})$ is the eigenvalue of band $\mu$. The interacting spin susceptibility and charge susceptibility in RPA level are given by,
\begin{eqnarray}
\chi^{RPA}_1(\bm{q})&=&[1-\chi_0(\bm{q})U^s(\bm{q})]^{-1}\chi_0(\bm{q}),\\
\chi^{RPA}_0(\bm{q})&=&[1+\chi_0(\bm{q})U^c(\bm{q})]^{-1}\chi_0(\bm{q}),
\label{RPA1}
\end{eqnarray}
where $U^s$, $U^c$ are the interaction matrices are,
\begin{eqnarray}
U^s_{\alpha l_1,\alpha l_2,\alpha l_3,\alpha l_4}(\bm{q})&=&
\begin{cases}
U_\alpha   & l_1=l_2=l_3=l_4,\\
U'_\alpha    & l_1=l_3\neq l_2=l_4,\\
J_\alpha   & l_1=l_2\neq l_3=l_4,\\
J'_\alpha   & l_1=l_4\neq l_2=l_3,\\
\end{cases}\\
\label{EQ:Us}
U^c_{\alpha l_1,\alpha l_2; \alpha l_3,\alpha l_4}(\bm{q})&=&
\begin{cases}
U_\alpha   & l_1=l_2=l_3=l_4,\\
-U'_\alpha+2J_\alpha   & l_1=l_3\neq l_2=l_4,\\
2U'_\alpha-J_\alpha   & l_1=l_2\neq l_3=l_4,\\
J'_\alpha   & l_1=l_4\neq l_2=l_3,\\
\end{cases}
\label{EQ:Uc1}
\end{eqnarray}
Here $\alpha$ is the sublattice index. For Nd site with $\alpha=A$, the interactions parameters are: $U_A=U$, $U'_A=U'$, $J_A=J$ and $J'_A=J'$. For Ni site with $\alpha=B$, the only non-vanishing interaction parameter is $U_{B}=U_3$. We plot the susceptibility in the main text, which is defined as $\chi_{0/RPA}=\frac{1}{2}\sum_{l_1,l_2}\chi^{0/RPA}_{l_1l_1;l_2l_2}(\bm{q},0)$. We also calculate the largest eigenvalues of the susceptibility matrix in momentum space (not shown), which is very similar to $\chi_{0}$.  The effective interaction obtained in the RPA approximation is,
\begin{eqnarray} V_{eff}=\sum_{ij,\textbf{k}\textbf{k}'}\Gamma_{ij}(\textbf{k},\textbf{k}')c^{\dag}_{i\textbf{k}\uparrow}c^{\dag}_{i-\textbf{k}\downarrow}c_{j-\textbf{k}'\downarrow}c_{j\textbf{k}'\uparrow}
\end{eqnarray}
where the momenta $\textbf{k}$ and $\textbf{k}'$ are restricted to  different FS $C_i$ with $\textbf{k}\in C_i$ and $\textbf{k}'\in C_j$ and $\Gamma_{ij}(\textbf{k},\textbf{k}')$ is the pairing scattering vertex in the singlet channel\cite{Kemper2010}. The pairing vertex is,
\begin{eqnarray}
\Gamma_{ij}(\textbf{k},\textbf{k}')=\sum_{l_1 l_2 l_3 l4}a^{l_2,*}_{v_i}(\textbf{k}) a^{l_3,*}_{v_i}(-\textbf{k}) Re[\Gamma_{l_1 l_2 l_3 l_4}(\textbf{k},\textbf{k}',\omega=0)] a^{l_1}_{v_j}(\textbf{k}') a^{l_4}_{v_j}(-\textbf{k}'),
\end{eqnarray}
 where $a^{l}_{v}$(orbital index $l$ and band index $v$) is the component of the eigenvectors from the diagonalization of the tight binding Hamiltonian. The orbital vertex function $\Gamma_{l_1 l_2 l_3 l_4}$ for the singlet channel and triplet channel in the fluctuation exchange formulation\cite{Bickers1989,Kubo2007,Kemper2010,Wu2014,Wu2015} are given by,
 \begin{eqnarray}
\Gamma^S_{l_1 l_2 l_3 l_4}(\textbf{k},\textbf{k}',\omega)&=&[\frac{3}{2}\bar{U}^s \chi^{RPA}_1(\textbf{k}-\textbf{k}',\omega)\bar{U}^s+\frac{1}{2}\bar{U}^s -\frac{1}{2}\bar{U}^c\chi^{RPA}_0(\textbf{k}-\textbf{k}',\omega)\bar{U}^c+\frac{1}{2}\bar{U}^c]_{l_1 l_2 l_3 l_4},\\
\Gamma^T_{l_1 l_2 l_3 l_4}(\textbf{k},\textbf{k}',\omega)&=&[-\frac{1}{2}\bar{U}^s \chi^{RPA}_1(\textbf{k}-\textbf{k}',\omega)\bar{U}^s+\frac{1}{2}\bar{U}^s -\frac{1}{2}\bar{U}^c\chi^{RPA}_0(\textbf{k}-\textbf{k}',\omega)\bar{U}^c+\frac{1}{2}\bar{U}^c]_{l_1 l_2 l_3 l_4},
\end{eqnarray}
where $\bar{U}^{s/c}=U^{s/c}(\bm{k}-\bm{k}')$.
The $\chi^{RPA}_0$ describes the charge fluctuation contribution and the $\chi^{RPA}_1$ the spin fluctuation contribution. For a given gap function $\Delta(\textbf{k})$, the pairing strength functional is,
\begin{eqnarray}
\lambda[\Delta(\textbf{k})]=-\frac{\sum_{ij}\oint_{C_i} \frac{dk_{\|}}{v_F(\textbf{k})} \oint_{C_j} \frac{dk'_{\|}}{v_F(\textbf{k}')} \Delta(\textbf{k})\Gamma_{ij}(\textbf{k},\textbf{k}') \Delta(\textbf{k}')} {V_G\sum_i\oint_{C_i} \frac{dk_{\|}}{v_F(\textbf{k})} [\Delta(\textbf{k})]^2 },
\label{strength}
\end{eqnarray}
where $v_F(\textbf{k})=|\triangledown_{\textbf{k}}E_i(\textbf{k})|$ is the Fermi velocity on a given fermi surface sheet $C_i$ and $V_G$ is the volume of the Brillouin zone. From the stationary condition we find the following eigenvalue problem,
\begin{eqnarray}
-\sum_{j} \oint_{C_j} \frac{dk'_{\|}}{V_G v_F(\textbf{k}')} \Gamma_{ij}(\textbf{k},\textbf{k}') \Delta_{\alpha}(\textbf{k}')=\lambda_{\alpha}\Delta_{\alpha}(\textbf{k}),
\label{strength1}
\end{eqnarray}
where the interaction $\Gamma_{ij}$ is the symmetric (antisymmetric) part of the full interaction in singlet (triplet) channel. The leading eigenfunction $\Delta_{\alpha}(\bf{k})$ and eigenvalue $\lambda_{\alpha}$ are obtained from the above equation. In the calculation, we treat those $k$ points, whose energies lie within a small energy window $\Delta E$ around the Fermi level, as effective $k$ points in the paring vertex function. We have checked the convergence of $\lambda$ with respect to $k$-mesh and $\Delta E$ ( with denser $50\times50\times30$ $k$ mesh and $\Delta E=0.01$ eV).

\section{pairing from the t-J model}

In the strong-coupling limit, similar to cuprate, we consider the inplane and outplane antiferromagnetic couplings between the spin of Ni $d_{x^2-y^2}$ orbital,
\begin{eqnarray}
H_{J}=\sum_{\langle ij\rangle}J_{ij}(\mathbf{S}_{i3}\mathbf{S}_{j3}-\frac{1}{4}n_{i3}n_{j3})
\end{eqnarray}
where
$\bm{S}_{i3}=\frac{1}{2}c_{i3\sigma}^{\dagger}\bm{\sigma}_{\sigma\sigma'}c_{i3\sigma'}$
is the local spin operator and $n_{i3}$is the local density
operator  for Ni $d_{x^2-y^2}$ orbital. $\langle ij\rangle$ denotes the inplane and outplane nearest neighbor(NN). The inplane coupling is $J_x=J_y=J_1$ and the outplane coupling is $J_z$. By performing the Fourier transformation, $H_{J}$ in momentum space reads
\begin{eqnarray}
H_{J}=\sum_{\mathbf{k,k}'}V_{\mathbf{k},\mathbf{k}'}c_{\bm{k}3\uparrow}^{\dagger}c_{-\bm{k}3\downarrow}^{\dagger}c_{-\bm{k}'3\downarrow}
c_{\bm{k}'3\uparrow},
\end{eqnarray}
with
$V_{\mathbf{k},\mathbf{k}'}=-\frac{2J_1}{N}\sum_{\pm}(cosk_x\pm cosk_y)(cosk'_x\pm cosk'_y)-\frac{2J_2}{N}cosk_zcosk'_z$. Here we investigate the pairing state for doped system and neglect the
no-double-occupance constraint on this $t-J$ model and perform a mean-field decoupling, similar to the iron based superconductors\cite{Seo2008}. With this, the total Hamiltonian can be written as,
\begin{eqnarray} H_{MF}&=&\sum_{\mathbf{k}}\Psi_{\mathbf{k}}^{\dagger}A(\mathbf{k})\Psi_{\mathbf{k}}+\frac{N}{2J_1}\sum_{\nu=s,d}|\Delta_{\nu}|^2
+\frac{N}{2J_2}|\Delta_{z}|^2,\\
A(\mathbf{k})&=&\left(\begin{array}{cc}
h(\mathbf{k}) & \Delta_{\uparrow\downarrow}(\mathbf{k}) \\
 \Delta^{\dagger}_{\uparrow\downarrow}(\mathbf{k}) & -h^{*}(-\mathbf{k}) \\
 \end{array}\right), \nonumber\\
 \Delta_{\uparrow\downarrow}(\mathbf{k})&=&\left(\begin{array}{ccc}
0 &  &  \\
  & 0 & \\
  &   &  \Delta_{3}(\mathbf{k})  \\
 \end{array}\right),
\end{eqnarray}
where $\Psi_{\mathbf{k}}^{\dagger}=(\psi^\dag_{\bm{k}\uparrow},\psi^T_{-\bm{k}\downarrow})$, $\Delta_{3}(\mathbf{k})  =  \Delta_{s}(cosk_x+cosk_y)+\Delta_{d}(cosk_x+cosk_y)+\Delta_zcosk_z$, and
\begin{eqnarray}
\Delta_{s/d} &=& -\frac{2J_1}{N}\sum_{\mathbf{k}'}d_{\mathbf{k}'\uparrow}(cosk'_x\pm cosk'_y),\\
\Delta_{z} &=& -\frac{2J_2}{N}\sum_{\mathbf{k}'}d_{\mathbf{k}'\uparrow}cosk'_z,
\end{eqnarray}
with $d_{\mathbf{k}'\uparrow}=\left\langle
c_{-\mathbf{k}'3\downarrow}c_{\mathbf{k}'3\uparrow}\right\rangle
$. $A(\mathbf{k})$ can be diagonalized by an unitary transformation
$U_{\mathbf{k}}$ with and the Bogoliubov quasiparticle eigenvalues $E_{m+3}=-E_{m}$ with $m=1,2,3$. The self-consistent gap equations are
\begin{eqnarray}
\Delta_{s/d} & = & -\frac{2J_1}{N}\sum_{\mathbf{k},m}(cosk_x\pm cosk_y)U_{6,m}^{*}(\mathbf{k})U_{3,m}(\mathbf{k})F[E_{m}(\mathbf{k})]\nonumber\\
\Delta_{z} & = & -\frac{2J_2}{N}\sum_{\mathbf{k},m}cosk_zU_{6,m}^{*}(\mathbf{k})U_{3,m}(\mathbf{k})F[E_{m}(\mathbf{k})]
\end{eqnarray}
where $F[E]$ is Fermi-Dirac distribution function, $F[E]=1/(1+e^{E/k_{B}T})$. The above equations can be solved self-consistently, varying the doping and the value of $J_1,J_2$. For $n=1.0$ and $n=0.8$, the obtained  $d_{x^2-y^2}$ gap and ground-state energy as a function of $J_1$ ($J_1=J_2$) are given in Fig.\ref{tjej} (a) and (b). With $J_1$ being larger than 0.05, the $d_{x^2-y^2} $superconducting gap and ground-state energy increases abruptly. With $J_1=J_2=0.1$, Fig.4 in the main text shows the superconducting gap a function of doping. We find that electron doping will significantly suppress superconductivity and the gap reaches the maximum with 0.1 hole doping. Further hole doping will suppress the gap size. The 3D gap function of $d_{x^2-y^2}$-wave pairing is shown in Fig.4 (b) for $n=0.8$, where the gap on the spherical Fermi surfaces almost vanish. These are consistent with the RPA calculations.

\begin{figure}[tb]
\centerline{\includegraphics[width=0.6\columnwidth]{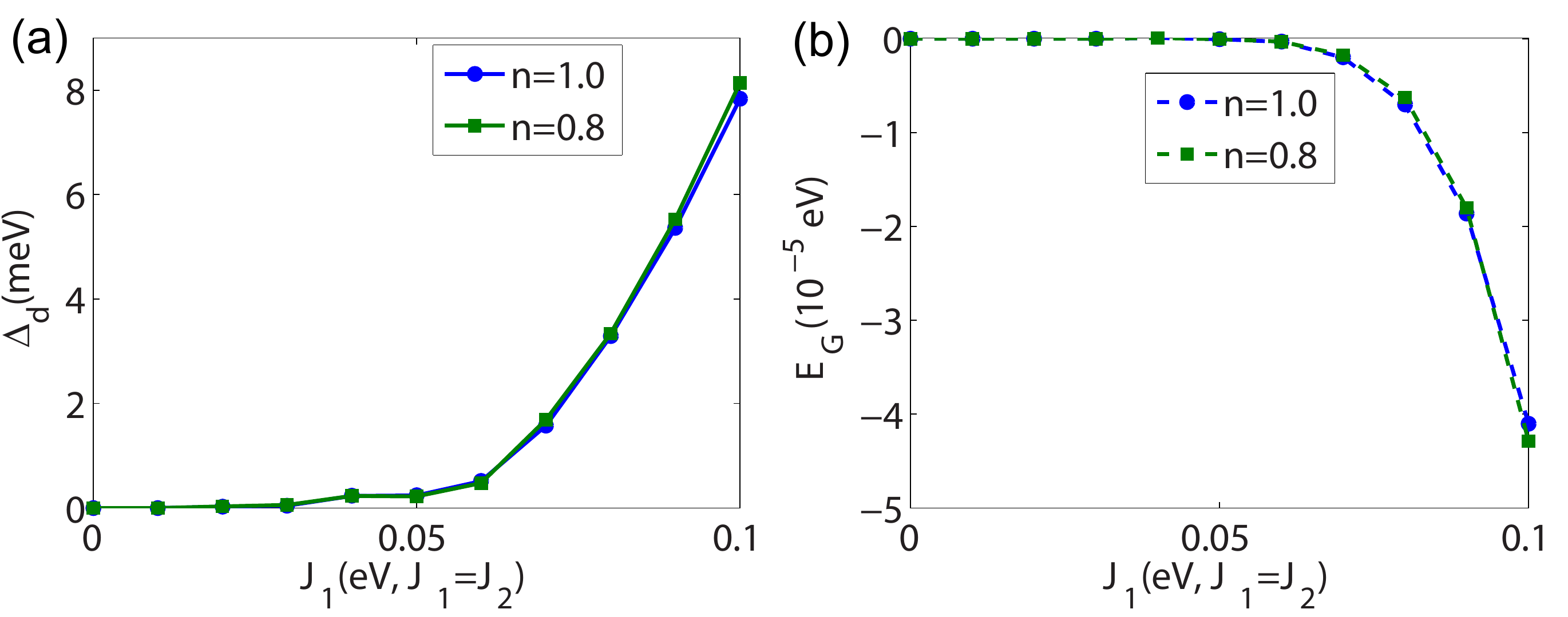}}
\caption{ (a) The gap function of $d_{x^2-y^2}$ pairing state as a function of $J_1$ for $n=1.0$ and $n=0.8$. (b) The ground-state energy relative to the normal state as a function of $J_1$ for $n=1.0$ and $n=0.8$. In the calculations, a $k$ mesh of $100\times100\times50$ is adopted.}
\label{tjej}
\end{figure}

\end{widetext}
\end{document}